\documentclass[preprint,showpacs,preprintnumbers,amsmath,amssymb]{revtex4}

\usepackage{amssymb}
\usepackage{graphicx}
\usepackage{dcolumn}
\usepackage{bm}

\newcommand{\EE}[1]{\mathbb{E}\left[#1\right]}
\newcommand{\OO}[1]{O\left(#1\right)}
\newcommand{\PP}[1]{\mathbb{P}\left[#1\right]}

\begin{document}

\preprint{APS/123-QED}

\title{Extreme values and fat tails of multifractal fluctuations}

\author{J.F. Muzy}
\email{muzy@univ-corse.fr}
\affiliation{SPE UMR 6134, CNRS, Universit\'e de Corse, 20250 Corte, France}
\author{E. Bacry}
 \email{emmanuel.bacry@polytechnique.fr}
\author{A. Kozhemyak}
\email{alexey@cmapx.polytechnique.fr}
\affiliation{CMAP, Ecole Polytechnique, 91128 Palaiseau, France}

\date{\today}

\begin{abstract}
In this paper we discuss the problem of the estimation of 
extreme event occurrence probability for data drawn from some 
multifractal process. We also study 
the heavy (power-law) tail behavior of probability density function 
associated with such data. 
We show that because of strong correlations, standard extreme value approach
is not valid and classical tail exponent estimators should be interpreted 
cautiously. Extreme statistics associated with multifractal
random processes turn out to be characterized by non self-averaging properties.
Our considerations rely upon some analogy between random
multiplicative cascades and the physics of disordered systems
and also on recent mathematical results about the so-called 
multifractal formalism.
Applied to financial time series, our 
findings allow us to propose an unified framemork
that accounts for the observed
multiscaling properties of return fluctuations, the volatility
clustering phenomenon and the observed ``inverse cubic law'' of the 
return pdf tails.
\end{abstract}

\maketitle

\section{Introduction}
\label{intro}
Statistics of extremes is an issue of prime importance in many situations
where extreme events may appear to have disastrous effects 
or to govern the main observations. Such situations can be found in a
wide  range of fields from physics (e.g. disordered
systems at low temperature), geology (earthquakes),
meteorology (rainfalls, storms), 
insurance or finance \cite{EVB1,EVB2,BSor}. 
Extreme events are particularily relevant for random phenomena
involving  a probability density
function (pdf) which tails decrease very slowly and roughly follow a
power-law. Such heavy tailed distributions have
been observed in many natural phenomena. An important question
concerns therefore the estimation and interpretation 
of pdf tail exponent as well as
the identification of mechanisms leading to them.
These problems are at the heart of an increasing
number of works \cite{BSor,BouPL01,New04}.   
Probabilistic and statistical questions related to very high or very
low values of random variables are addressed within the framework of
extreme value theory. This theory has been originally developed 
for independent identically distributed (i.i.d.)
random variables and more recently extended to stationary processes
where independence condition has been relaxed \cite{EVB1}. 
However, when correlations are not weak enough very few results 
are known. 

In this paper we aim at studying the statistics of extreme events
and the (fat) tail exponent of fluctuations associated with multifractal 
random processes. Mutifractal extreme fluctuations are interesting because
they represent an example of strongly correlated random variables that 
do not satisfy standard mixing conditions of extreme value theory.
But multifractals are also interesting because they are widely used to model of
self-similar phenomena displaying multiscaling properties.
Our purpose is to study probability of extreme event of multifractal processes and 
the associated power-law tail exponent.
We will show that for such processes,
the pdf tail exponent value observed (estimated) from experimental data 
may be different from the value associated with the unconditional 
theoretical pdf. 
We examine different experimental 
conditions depending upon the size of the observed sample $L$, 
the correlation length $T$ and the observation scale $\tau$. We 
emphasize that, under most usual conditions, the
estimated tail exponent is smaller than the exponent one would expect 
without correlations. This result is intimately related
to some non-self averaging property of usual tail exponent estimators
and is the analog of the glassy behavior observed at low
temperatures in disordered systems. 
We apply our phenomenological framework to multifractal models
of asset return fluctuations and show that the well known
``inverse cubic law'' of pdf tails can be naturally explained in terms
of volatility correlations.

The paper is organized as follows: In section \ref{sec_ev}, we briefly
review the main results about extreme value theory and the commonly
used tail exponent estimators. Multifractal processes, multiplicative
cascades and their
main mathemetical properties are reviewed in section \ref{sec_mf}.
In sections \ref{sec_mfev} we build an extreme value theory
for multifractal cascades. The problem of the estimation of the
power-law tail exponent associated with the cumulative probability
distribution of multifractal fluctuations is addressed in section
\ref{sec_tail}. In section \ref{sec_num} we illustrate
our phenomenology by numerical examples of continuous as well
as discrete cascades.
Application to finance is considered in section \ref{sec_finance}
while section \ref{sec_conclusion} contains
concluding remarks and questions for future research.
Auxiliary computations or technical material are reported in Appendices.

\section{Fat tails and extreme value statistics}
\label{sec_ev}
Let us briefly review the main estimators used to 
characterize the power-law tail behavior of some probability
distribution. Let $F(x) = \PP{Z \leq x}$ be the cumulative probability
distribution (cdf) of some random variable $Z$. The variable $Z$ is said to be of
power-law type tail or Pareto type tail if, when $x \rightarrow +\infty$,
\begin{equation} 
\label{pareto}
1-F(x) \sim C x^{-\mu}
\end{equation}
where $C$ is a positive normalization constant or a slowly varying function. 
The exponent $\mu$ is called 
the {\em tail exponent} of the distribution. The problem addressed
in this section concerns the estimation of this exponent from
empirical data.

A simple, widely used method relies on the so-called ``Zipf'' or
``rank-frequency''plots (see e.g. \cite{Bman,BSor,New04}): Let
$Z_1 \ldots Z_N$ be $N$ i.i.d. samples characterized by the same
distibution function $F(x)$. Let us denote 
$X_1 \geq X_2 \geq \ldots \geq X_N$ the rank ordered values of $Z_i$ 
(sorted in descending order). 
Then, if the asymptotic behavior of $F(x)$ is Pareto like (as in Eq. (\ref{pareto})), 
one has, for $ 1 \leq j \leq k \ll N$:
\begin{equation}
\label{texpp}
  1-F(X_j) \sim \frac{j}{N} \Longrightarrow X_j \sim \left(\frac{j}{N}\right)^{-1/\mu}
\end{equation}
and therefore $\mu$ can be simply estimated as the slope of
the {\em Zipf plot} $\left[\ln X_j, \ln(j)\right]$, $j = 1 \ldots k$ (see \cite{Bman} for exact results). 
In the following, we refer to this estimator as the {\em power-law fit estimator}.
Since this estimator is biased \cite{New04}, one should use alternative estimators.

Actually, there are many alternative tail exponent 
estimators  \cite{EVB2}. The most commonly used are 
Hill or Pickands estimators that are defined as follows:
Let $k(N) = o(N)$ be the maximum $X$ rank used to estimate $\mu$, 
the Hill estimator is simply
\begin{equation}
  \mu_H(k,N) = \frac{k-1}{\sum_{i=1}^{k-1} \ln(X_i/X_k)}
\end{equation}
while the Pickands estimator is:
\begin{equation}
  \mu_P(k,N) = \frac{\ln(2)}{\ln\left(\frac{X_k-X_{2k}}{X_{2k}-X_{4k}}\right)}
\end{equation}

The mathematical study of these estimators (consistency, bias, asymptotic normality) 
relies upon Extreme Value Theory \cite{EVB1,EVB2}, i.e., the theory that deals with maxima and minima properties  of random variables.
According to this theory, the maximum value $Y$ of $N$ i.i.d. random variables (normalized properly), has a probability density 
function that asymptotically belongs to the Fisher-Tippett's Extreme
Value distribution class. According the the shape of the pdf of $Y$, the pdf of the maximum can either be of the Frechet type  ($ F_Y(x) = e^{-x^{-\mu}}$), 
of the Gumbel type ($ F_Y(x) = e^{-e^{-x}}$) or of the Weibull class \cite{EVB1}. 

Extreme value theory has also been extended to dependent (or correlated) 
stationary random processes \cite{EVB1}.
Under mixing conditions ensuring asymptotic 
independence of maxima, it can be shown that the limit theorems 
established in the i.i.d. case still hold.
 The key difference is that the number $N$ of independent variables is replaced by an ``effective''
number $N \theta$. The value $0 < \theta < 1$ that 
quantifies the effect of dependence is called the {\em extremal index} \cite{EVB1}.
In the case of  Gaussian processes, these theorems hold
provided the covariance function $\rho(x)$
decreases sufficiently fast for large lags $x$, i.e.
\begin{equation}
\label{mixingC}
  \rho(x) \ln(x) \operatornamewithlimits{\rightarrow}_{x \rightarrow +\infty} 0
\end{equation}
A simple intuitive justification of $1/\ln(x)$ as the limiting case for
the validity of standard extreme value theorems is provided in ref. \cite{CarLeDou01}.

The main purpose of this paper is to try to understand how these theoretical results apply
to data sampled from a multifractal process.

\section{Multifractal processes}
\label{sec_mf}
Multifractal processes are random functions that possess 
non trivial scaling properties.
They are now widely used models in many areas
of applied and fundamental fields. Well known examples 
are turbulence, internet traffic, rainfall distributions or finance. 
For the sake of simplicity we will consider only non-decreasing multifractal 
processes (often referred to as  multifractal {\em measures} though their variations are not bounded) denoted hereafter $M(t)$.
More general multifractal processes can be conveniently builded
as a simple Brownian motion $B(t)$ compounded with the measure $M(t)$ considered
as a stochastic time: $X(t) = B\left[M(t)\right]$
The statistical properties of $X(t)$ can be directly deduced from those of $M(t)$ (see e.g., \cite{MuBa02,BaMu03}).

\subsection{Multiscaling}
Multifractal processes are characterized by  multiscaling
properties of their variations. More precisely, if one defines the increments of $M(t)$ 
at scale $\tau$, $M(t,\tau) = M(t+\tau)-M(t)$,
multifractality can be loosely defined from the scaling behavior of the moments
of $M(t,\tau)$:
\begin{equation}
\label{gscaling}
  \EE{M(t,\tau)^q)} \operatornamewithlimits{\sim}_{\tau \rightarrow 0} \tau^{\zeta(q)} 
\end{equation}
where $q \in \mathbb{R} $ is the order of the moment and
the exponent $\zeta(q)$ is some nonlinear convex function often
called the multifractal exponent spectrum.
The simplest example of such function is the so-called
{\em log-normal} spectrum for which $\zeta(q)$ 
is a simple parabola:
\begin{equation}
\label{ln}
  \zeta(q) = (1+\frac{\lambda^2}{2}) q -\frac{\lambda^2}{2} q^2
\end{equation}
The coefficient $\lambda^2$ quantifies the curvature of $\zeta(q)$
(and hence the multifractality of the process) 
that is constant in the log-normal case.
In the general case, one often calls $-\zeta''(0)$
the {\em intermittency coefficient}.
Let us notice that, because of H\"older inequality for moments the
scaling (\ref{gscaling}) with a non linear convex $\zeta(q)$ cannot hold
at arbitrary large scales $\tau$ but is valid only in a domain bounded by
some large scale $T$ (actually the limit $\tau \rightarrow 0$ in 
Eq. (\ref{gscaling}) must be understood as $\tau/T \ll 1$). This scale $T$ will be called the {\em integral scale}.

\subsection{Singularity spectrum and multifractal formalism}
\label{mf}
Let us recall some classical results about the multifractal
formalism. This formalism has been introduced in
early eighties by Parisi and Frisch (see e.g. \cite{BF,HJKP}) in order to interpret 
the above multiscaling properties of the moments 
in terms of pointwise regularity properties of the paths of the process $M(t)$.
Let us define the local
H\"older exponent $\alpha(t_0)$ at point (or time) $t_0$ as
\begin{equation}
   M(t_0,\tau) \operatornamewithlimits{\sim}_{\tau \rightarrow
   0} \tau^{\alpha(t_0)}
\end{equation}
The limit $\tau \rightarrow 0$ means $\tau \ll T$ where $T$ is 
the integral scale.
The singularity spectrum $f^\star(\alpha)$ can be introduced as the fractal
(Haussdorf or packing) dimension of the iso-H\"older exponents sets:
\begin{equation}
  f^\star(\alpha) = Dim \{ t, \alpha(t) = \alpha \}
\end{equation}
Roughly speaking, this equation means that at scales $\tau \ll T$, the number of
points where $M(t,\tau) \sim \tau^\alpha$ is 
\begin{equation}
N(\tau,\alpha) \sim \tau^{-f^\star(\alpha)} \; .
\end{equation}

The {\em multifractal formalism} states that $f^\star(\alpha)$ and $\zeta(q)$ as defined
in Eq. (\ref{gscaling}) are basically
Legendre transform one to each other. More precisely, if we define $f(\alpha)$ as the Legendre transform of $\zeta(q)$, i.e.,
\begin{eqnarray*}
   f(\alpha) & = & 1+\min_q(q\alpha-\zeta(q)) \\
   \zeta(q) & = & 1 + \min_\alpha(q\alpha-f(\alpha)),
\end{eqnarray*} 
then 
\[
f^\star(\alpha) = f(\alpha),~~~\forall \alpha^{\star} \ge \alpha \ge \alpha_{\star},
\]
where $\alpha_\star$ and $\alpha^{\star}$ are defined by
\begin{eqnarray*}
   \alpha_\star  & = & \inf\{\alpha,~f^\star(\alpha) = 0\} \\
   \alpha^\star  & = & \sup\{\alpha,~f^\star(\alpha) = 0\}. 
 \end{eqnarray*}

In the following sections, we will use the fact that $q$ can be interpreted as a 
value of the derivative of  $f(\alpha)$ and conversely
$\alpha$ is a value of the derivative of $\zeta(q)$ :
for a given value of $q = q_{0}$ one has, from previous Legendre
transform relationship and thanks to the convexity of $\zeta(q)$:
\begin{eqnarray}
\label{yes}
 f(\alpha_{0}) &= &1 + q_{0}\alpha_{0} -\zeta(q_{0}) \\
\alpha_{0} & = & \frac{d \zeta}{dq} (q_{0})\\
 q_{0} & = & \frac{d f}{d\alpha}(\alpha_{0}).
 \end{eqnarray}

Let us note that, $f^\star(\alpha)$ can be seen as the Legendre transform of the function $\zeta^\star(q)$  simply defined as
\begin{equation}
\label{zetastar}
\zeta^\star(q) = 
\left\{
\begin{array}{ll} 
    \zeta(q) & {\mbox{for}}  ~~q^{\star} \leq q \leq q_{\star} \\
     \alpha_\star q & {\mbox{for}}  ~~q > q_{\star}, \\
      \alpha^\star q & {\mbox{for}}  ~~q < q^{\star}, 
\end{array}
\right.,
\end{equation}
where 
\begin{eqnarray}
\label{qstar}
q_{\star} & = & \frac{df}{d\alpha}(\alpha_{\star}) \\
q^{\star} & = & \frac{df}{d\alpha}(\alpha^{\star})
\end{eqnarray}

It is important to point out that, experimentally, under usual conditions, only $\zeta^{\star}(q)$ (and not $\zeta(q)$) can be estimated 
(see e.g. refs. \cite{Mol96,LAC04,MuBa05}).

\subsection{Cascades}
\label{sec_casc}
The paradigm of multifractal measures
are multiplicative cascades originally 
introduced by the russian school for modelling the energy 
cascade in fully developed turbulence. After the early works of
Mandelbrot \cite{Man74a,Man74b,Man03}, a lot of mathematical studies have been devoted to discrete  
random cascades \cite{KaPe76,Gui87,Mol96,Mol97,Liu02}. 
Very recently, continuous versions of these cascades have been
defined: they share exact multifractal scaling with discrete cascades
but they display continuous scaling and possess stationary
increments \cite{MuDeBa00,BaMan02,MuBa02,BaMu03}.  
Let us summary the main properties of these constructions
and set some notations.

The simplest discrete multifractal cascade can be constructed as follows:
one starts with an interval of length $T$ where the measure is
uniformly spread, and split the interval in two equal parts: On each 
part, the density is multiplied by (positive) i.i.d. random factors $W$.
Each of the two sub-intervals is again cut in two equal parts and
the process is repeated infinitely. At construction step $n$, 
if one addresses a dyadic interval of length $T2^{-n}$ by a
kneading sequence $k_1 \ldots k_n$, with $k_i = 0, 1$,
the ``mass'' of this interval (denoted as $I_{k_1 \ldots k_n}$) 
is simply:
\begin{equation}
\label{consM}
  M_n (I_{k_1 \ldots k_n}) = 2^{-n} \prod_{i=1}^{n} W_{k_1 \ldots k_i}
  = 2^{-n} e^{\sum_{i=1}^n \omega_{k_1 \ldots k_i}}
\end{equation} 
where all the $W_{k_1 \ldots k_i} = e^{\omega_{k_1 \ldots k_i}}$ are i.i.d such that $\EE{W} = 1$.
Peyri\`ere and Kahane \cite{KaPe76} proved that
this construction converges almost surely towards a stochastic non decreasing process
$M_{\infty}$ provided $\EE{W \ln W} < 1$.
The multifractality of $M_{\infty}$ (hereafter simply denoted as $M$) 
and the validity of the previously described multifractal formalism have been studied
by many authors (see e.g. \cite{Mol96}). 
An interesting additional property of cascades is that they are self-similar in the 
following stochastic sense:
\begin{equation}
\label{ssM}
  M [I_{k_1 \ldots k_n}] \operatornamewithlimits{=}_{law} 2^{-1} W M [I_{k_1 \ldots k_{n-1}}]
\end{equation}
and therefore the order $q$ moments of $M(t,\tau)$ behave as a power-law:
\begin{equation}
\label{sscasc}
  \EE{M[0,T2^{-n}]^q} = 2^{-nq} \EE{W^q}^n \EE{M[0,T]^q} 
\end{equation}
Comparison of Eqs. (\ref{sscasc}) and (\ref{gscaling}) with $\tau = T2^{-n}$ directly
yields the expression of the spectrum $\zeta(q)$ in terms of cumulant generating
function of $\omega = \ln W$:
\begin{equation}
\label{zetaM}
  \zeta(q) = q-\ln_2(\EE{W^q})
\end{equation}
In that sense, $\zeta(q)$ is nothing but a {\em large deviation} spectrum.
Let us mention that the validity of the multifractal formalism has been 
rigorously proved for cascades.
In Appendix \ref{ap_excasc} we provide explicit expressions of
$\zeta(q)$ for various laws of $W$.

Let us note that, as shown in \cite{AMS98,ABMM98}, the correlation function $\rho(x)$ of such cascades decreases slowly as 
\begin{equation}
\label{correlation}
\rho(x) \simeq \ln(T/x) ~~{\mbox{for lags}}~ x \leq T.
\end{equation} 
If we consider that the data come from a sampling  of successive independent cascades, the correlation function is 0 for lags above $T$, i.e.,
\begin{equation}
\label{correlation1}
\rho(x) = 0 \; ,~~x \ge T.
\end{equation} 
The integral scale $T$ where cascading process ``starts'' can therefore be interpreted 
as a {\em correlation length} for the variations of the $M(t)$.  

Because the previous construction involves dyadic intervals, 
and a 'top-bottom' construction,
it is far from being stationary. In order to get rid of this drawback,
as already mentionned, some continuous cascade constructions 
have been recently proposed and
studied on a mathematical ground \cite{MuDeBa00,BaMan02,MuBa02,BaMu03}. 
Without entering into details, we just want to 
mention that such continuous cascades
involve a family of infinitely divisible random processes $\omega_l(t)$ which
correlation function basically follows Eqs (\ref{correlation}) 
and (\ref{correlation1}).
The process $e^{\omega_l(t)}$ is the analog of the density 
satisfying the self-similarity:
\[
e^{\omega_{sl}(st)} \operatornamewithlimits{=}_{law} e^{\Omega_s} e^{\omega_l(t)} 
\]
Martingale theory allows one to prove the convergence of 
the continuous process $M[0,t] = \lim_{l \rightarrow 0} \int_0^t e^{\omega_l(u)} \; du$.

Let us mention that the validity of the multifractal formalism has been
established by Molchan by for discrete cascades
\cite{Mol96,Mol97} and generalized by Barral and Mandelbrot  
for continuous cascades \cite{BaMan02}.

In the sequel, we will be using indifferently the classical top-bottom model (i.e., we will consider that the data come from a sampling of a succession of independent realizations of the same cascade process) and the continuous cascade model (i.e., we will consider that the data simply come from a sampling of a continuous cascade). Most of the arguments will be done using the first model while numerical examples will be performed on data of the second model.

\subsection{Cascades have fat tails}
Let us first emphasize that the unconditional law of $M$ can have a power law 
tail. Indeed, a simple argument involving the self-similarity of the 
limit measure allows one to obtain a simple bound: since the 
measure is additive, $M[0,1] = M[0,1/2]+M[1/2,1]$,
for $q>1$ one gets
\begin{equation}
   \EE{M[0,1]^q} \geq 2 \EE{M[0,1/2]^q} = 2^{1-\zeta(q)} \EE{M[0,1]^q}
\end{equation}
and finally $\EE{M[0,1]^q} < +\infty \Rightarrow \zeta(q) \geq 1$.
The reverse implication is basically true. It is however trickier to obtain and we refer to \cite{KaPe76}
for a precise proof.

Since the power law tail exponent of a distribution is directly related the the maximum order 
finite moments, if one defines
\begin{equation}
\label{defmu} 
 \mu = 1+\sup\{q, q > 1, \zeta(q) > 1 \} \; , 
\end{equation}
then , cascades  (discrete as well as
continuous) have thick tails of exponent $\mu$: 
\begin{equation}
\label{texp}
 \PP{M(t,\tau) \geq x} 
\operatornamewithlimits{\sim}_{x \rightarrow +\infty} x^{-\mu} \; .
\end{equation}

\subsection{Defining the asymptotic limit $N \rightarrow +\infty$}
\label{sec_mix}
 As we have seen (Eq. (\ref{correlation})), the covariance function $\rho(x)$ for multifractal processes has a very slow decay (slower than Eq. (\ref{mixingC})) up to the {\em integral scale} $T$ above which data are independent.
Thus we expect that mixing conditions are not valid in the range from the sampling scale $\tau$ to the integral scale $T$ while they hold 
when ``looking'' above scale $T$. Let us try to be more precise: the mixing conditions are conditions in the limit when the total 
number of samples $N$ goes to $+\infty$. However, there are several ways 
to reach this asymptotic limit.
Let $L$ be the length of the whole sequence and $\tau$ the sampling scale. The total number of samples is therefore,
\[
N = \frac L \tau,
\]
while the number of integral scales $N_{T}$ and the number of samples per integral scales $N_{\tau}$ are
respectively:
\begin{equation}
\label{Ntau}
N_{T} = \frac L T~~~,~~N_{\tau}=\frac T  \tau,~~N = N_{\tau}N_{T}.
\end{equation}
In order to control the relative values of $N_T$ and $N_\tau$, let us 
define the exponent $\chi$ as follows:
\begin{equation}
\label{defchi}
N_{T}\sim N_{\tau}^{\chi}
\end{equation}
Let us note that this exponent $\chi$ has been already introduced by
B.B. Mandelbrot as an ``embedding dimension'' \cite{Man90}
in order to discuss the concept of negative dimension (see below).

Thus,  when $N \rightarrow +\infty$, if, for instance, $\chi = 0$, it means that we are in the case $\tau \rightarrow 0$ (while $L$ and $T$ are fixed), i.e., most of the data are lying between the lags $\tau$
 and $T$. Consequently, we do not expect the mixing conditions to hold. On the contrary, if $\chi = +\infty$, it means that we are in the case $L\rightarrow +\infty$ (while $\tau$ and $T$ are fixed), i.e,  
most of the data are lying between the lags $T$ and $L$ and consequently the mixing conditions are satisfied. 
 
Thus, as it will be discussed in the next two sections,  in the first case ($\chi = 0$), nothing guarantees that classical 
results of extreme value theory can be applied and that exponent tail estimators provide the expected values,
whereas in the second case ($\chi = +\infty$), we expect the i.i.d. extreme value theorems 
to hold and the exponent tail estimators to be consistent.
As we will see, one can go continuously from the first case to the second one.
Actually we will show that both the extreme value distribution 
associated with cascades and the corresponding tail exponent estimator strongly depend on the value of $\chi$.

\section{Multifractal extreme value statistics}
\label{sec_mfev}
\subsection{Cumulative probability distribution of the maximum}
\label{sec_cum}
Let $\tau = T2^{-n}$. We call $X_{1}(N)$ the maximum value of $\ln(M(I_{n})/\tau)$, 
where $I_n$ is a short notation of the dyadic intervals $I_{k_1 \ldots k_n}$ of size $\tau$. Let 
$P(x,N) = \PP{X_1(N) < x}$ be the cumulative distribution function (cdf) of $X_1(N)$, i.e., the probability that $X_1(N)$
is smaller than $x$. 

Let us recall that we consider that the data come from a sampling of successive independent realizations of the same cascade measure. 
Thus $T$ (the integral scale) is fixed whereas $L$ (the total length of the data) and $\tau$ (the sampling scale) are varying. 
We want to study the statistics $P(x,N)$  of the data in the limit $N\rightarrow +\infty$. 
As we explained in the previous section, these statistics will strongly depend on $\chi$ (Eq. (\ref{defchi})), i.e., on the way $N_{\tau}$ and $N_{T}$ go to $+\infty$. We fix $\chi = r/p$ and we choose the following parametrization:
\[
N_{\tau} \sim 2^{pm},~~~N_{T} \sim 2^{rm},~~~N = N_{\tau}N_{T} \sim 2^{(p+r)m},
\]
when the integer parameter $m\rightarrow +\infty$. 

Thus $P(x,m) \equiv P(x,N(m))$ is the cdf of $\ln M(I_{pm})/\tau$ where $\tau = T2^{-pm}$.

In Appendix \ref{ap_iter}, we show that $P(x,m)$ 
can be simply expressed as: 
\begin{equation}
\label{iter}
  P(x,m) =  \left(P'(x,pm) \right)^{2^{rm}} 
\end{equation}
where $P'(x,n)$ is the cdf associated with the maximum of $\ln(M(I_{n})/\tau)$ on 
a single integral scale instead of the whole data. It satisfies the renormalization equation:
\begin{equation} 
\label{iter1}
   P'(x,n+1) = \left[P'(x,n) \star g(x)\right]^2
\end{equation}
where $\star$ stands for the convolution product and 
$g(x)$ is the probability density of $\omega = \ln(W)$. 

Let us notice that the initial condition $P(x,0)$ is precisely given by the 
law of $\ln(M)$ which exponential tail is described by Eq. (\ref{texp}): 
\begin{equation}
\label{inicond}
  P(x,0) \operatornamewithlimits{\sim}_{x \rightarrow +\infty} 1- Ce^{-\mu x}
\end{equation}

\subsection{Traveling front solutions}
\subsubsection{Case $\chi = +\infty$}
When $r=1$ and $p = 0$, Eq. (\ref{iter})
is simply the recurrence for the maximum cdf of i.i.d. random variables 
which solutions are Fisher-Tippett's
fixed points reviewed in sec. \ref{sec_ev}. More precisely, since the initial condition is 
exponentially descreasing (Eq. (\ref{inicond})), when $N \rightarrow +\infty$,
$P(x,N)$ will have a Gumbel shape. Consequently, the law of the maximum value of $M$ will belong
to the Frechet class with a tail exponent $\mu$ as defined in Eq. (\ref{defmu}).

\subsubsection{Eq. (\ref{iter1}) and the KPP equation}
In order to solve the nonlinear problem (\ref{iter}) in the general case, let
us first study Eq. (\ref{iter1}) and its solutions.
We will show that these solutions are traveling fronts so that, in some 
moving frame, Eq. (\ref{iter}) reduces to a standard (i.i.d.) extreme value problem.

Equation (\ref{iter1}) is exactly satisfied
by the cdf associated with the maximum value of random variables hierachically correlated 
(generated additively along a Cayley tree). Such an equation 
has been studied in ref. \cite{DerSpo88,DeMaj01}. In ref. \cite{CarLeDou01},
the authors have considered a log-normal random process with
log-correlated covariance, that can be considered as a continuous
version of a random cascade. 
In this case, the obtained partial differential 
equation for the law of the maximum turns out to be
the famous Kolmogorov-Petrovsky-Piscounov (KPP) equation. 
Thus Eq. (\ref{iter1}), in the case where $g(x)$ is Gaussian, 
can be seen as a ``discretized'' version of the KPP equation.

It is well know that the KPP equation has 
traveling wavefront solutions connecting 
the homegeneous stable state
to the unstable one. These solutions can be 
studied using linear stability analysis. 
As emphasized in \cite{VS03}, most of KPP features
are somehow universal in the sense that the same analysis 
can be performed for a wide variety
of ``reaction-diffusion'' problems involving non linear integro-differential or integro-diffence equations.
In particular the famous marginal stability criterium (see e.g. refs. 
\cite{VS03,Brunet,MajKra02} and below) 
for the selected front velocity can be generically applied for a large class of equations.
Therefore, as explained in \cite{Brunet,DeMaj01,MajKra02}, one can apply the same techniques 
to get solutions of Eq. (\ref{iter1}). Though these studies do not rely on a fully rigorous
mathematical ground (as KPP), we reproduce a similar analysis in the following.

Let us first notice that both Eqs (\ref{iter}) and Eqs (\ref{iter1}) admit two homogeneous
solutions $P = 0$ and $P = 1$, the first one being (linearly) stable while the second one is 
unstable. A given cdf $P(x,N)$ (or $P'(x,n)$) will therefore connect the
stable state to the unstable one. 
As in the above cited references, one can consider a traveling front solution of (\ref{iter1})
$P'(x,n) = P'_{tw}(x-v'n)$ 
where $v'$ is the front velocity.
In order to compute this velocity, one performs a linear analysis in the
vicinity of the unstable solution, i.e.,
in the limit $x\rightarrow +\infty$ where $P'_{tw}(x-v'n)\rightarrow 1$.

If we denote
\[
  Q_{tw}'(x) \equiv 1-P_{tw}'(x) \ll 1 \;
\]
then, to the first order in $Q_{tw}'(x)$, Eq. (\ref{iter1}) becomes :
\[
  Q_{tw}'(x-v') = 2 Q_{tw}'(x) \ast g(x) + \OO{Q_{tw}'^2}
\]
If one seeks for exponential solutions:
\[
  Q_{tw}'(x) = C e^{-qx} \; ;
\]
then,
\begin{equation}
\label{solpp}
  P_{tw}'(x,n) = 1-Ce^{-q(x-v'(q)n)} \; ,
\end{equation}
where $v'$ and $q$ satisfy the ``dispersion'' relation:
\[
  v'(q) = q^{-1} \ln \left[2 \int g(x)e^{q x} \; dx \right] 
\]
Notice that $g(x)$ is the law of the logarithms of the weights 
of the cascade construction, and then, from Eq. (\ref{zetaM})
\[
\ln( 2 \int g(x) e^{qx} \; dx )  = \ln(2)\left(1+q-\zeta(q)\right) 
\]
which yields:
\begin{equation}
\label{vp}
  v'(q) = \frac{\ln(2)\left(1+q-\zeta(q)\right)}{q} 
\end{equation}
In the next section, one shall see which velocity (or which $q$ value) is selected

\subsubsection{Case $\chi \neq +\infty$}
From Eqs. (\ref{iter}) and (\ref{solpp}), one gets the following traveling front solution for $P$,
\[
  P(x,m) = (1-Ce^{-q(x-v'(q)pm)})^{2^{rm}} \; ,
\]
where $v'(q)$ is given by (\ref{vp}).

If $y = x-v'(q)pm-q^{-1}r \ln(2) m$, the cdf of $y$ converges to the
Gumbel shape:
\[
  P(y,m) = (1-Ce^{-qy-r\ln(2)m})^{2^{rm}} \operatornamewithlimits{\longrightarrow}_{m \rightarrow +\infty} e^{-Ce^{-qy}}
\] 
and thus $P(x,m)$ is itself a traveling Gumbel front:
\begin{equation}
\label{gumbelfront}
  P(x,m) = e^{-Ce^{-q(x-m v(q))}}
\end{equation}
with the dispersion relation:
\begin{equation}
\label{drel}
  v(q) = \frac{\ln(2)\left(r+p(1+q-\zeta(q))\right)}{q} 
\end{equation}

One can reproduce the same kind of analysis as in 
refs \cite{Brunet,BruDer97,MajKra00,DeMaj01,MajKra02}. 
In Appendix \ref{ap_results}, we provide a sketch of proof that 
one can use a standard Aronson-Weinberger stability criterium 
to compute the velocity and $q$ value which are selected:
Let $q_{min}$ be the unique positive $q$ value such that
\begin{equation}
\label{awc}
  v(q_{min}) = \min_{q>0} v(q)
\end{equation}
and let 
\begin{equation}
\label{qstarchi}
q_{\star,\chi} = \min(\mu,q_{min}).
\end{equation}
Then the selected velocity is simply $v(q_{\star,\chi})$ and the shape of
the traveling front is (up to a subdominant correction described
in Appendix \ref{ap_results})  
\[
  P(x,m) = e^{-Ce^{-q_\star(x-v(q_{\star,\chi})m)}}
\] 
Let us recall (see section \ref{sec_cum}) that $P(x,m)$ corresponds to the cdf of 
$\ln M(I_{n})/\tau$ where $\tau = T2^{-n}$ and $n = pm$. Let us go back to $\ln M$.

\subsubsection{From $\ln M(I_n) / \tau$ to $\ln M(I_n)$}
The velocity $v(q)$ is a velocity related to the parameter $m$, i.e., 
$P(x,m)=P(x-v(q)m)$. It is convenient to compute 
the velocity $v_{\tau}(q)$ as respect to the ``observable'' scale
$\ln(T/\tau)$, i.e., $P(x,m)=P(x-v_{\tau}(q)\ln(T/\tau))$. Clearly, $v_{\tau}(q)=v(q)m/\ln(T/\tau)$. Since $\tau = T2^{-n}$ and $n=pm$, one gets
$v_{\tau}(q) = v(q)/p\ln(2)$.
Now, if we switch from the cdf of $\ln M(I_{n})/\tau$  to the cdf of 
$\ln M(I_{n})$, one finally obtains the velocity:
\[
  v_{\ln M}(q)  \equiv v(q)/p\ln(2) - 1
     =  -\frac{\zeta(q)-1-\chi}{q}
\]
Let us note that the minimum of $v_{\ln M }(q)$ (i.e., the minimum of $v(q)$) is reached for $q_{min}$ satisfying
\[
    -\chi  =  1+q\frac{d \zeta(q_{min})}{dq}-\zeta(q_{min}).
 \]
One can also notice that $1+q\frac{d \zeta(q)}{dq}-\zeta(q)$ is the Legendre transform of $\zeta(q)$. We know from section \ref{mf} that  this Legendre transform is nothing but the spectrum $f(\alpha)$. According to 
the results of this section (and particularly Eq. (\ref{yes})) 
it can be easily seen that 
this value of $q_{min}$ corresponds to a singularity 
exponent $\alpha_{min}$ with
\begin{eqnarray}
 f(\alpha_{min}) & = & -\chi \\
 q_{min} & = & \frac{df}{d\alpha} (\alpha_{min})\\
 \alpha_{min} & = & v_{\ln M}(q_{min}).
 \end{eqnarray}
Moreover, since $f(\alpha_{min}) \le 0$ then $df(\alpha_{min})/d\alpha>0$ and consequently $q_{min} > 0$. Thus,
 this value of $q_{min}$ does correspond to the one defined in Eq. (\ref{awc}).

We finally established that the law of the maximum of
$M$ at scale $\tau$ of a sample of length $\tau^{-\chi}$ is Frechet 
when $\tau \rightarrow 0$ with a tail exponent $q_{\star,\chi}$ that depends
on $\chi$:
\begin{eqnarray}
\label{tailmax}
   q_{\star,\chi} & = & \min(q_{min},\mu) \nonumber \\
   q_{min} & = & \frac{df}{d\alpha}(\alpha_{min}) \\
   f(\alpha_{min}) & = & -\chi \nonumber 
\end{eqnarray}

Let us remark that for $\chi = 0$, the value of $q_{\star,\chi}$ is exactly the value $q_{\star}$ that is involved in the multifractal formalism as described at the end of section \ref{mf} (Eq. (\ref{qstar})). For finite $\chi$ value, 
as already noticed more than a decade ago by Mandelbrot \cite{Man90},
the exponent $\chi$ that governs the size of the ``supersample'' 
allows one to explore negative dimensions (i.e., negative values of $f(\alpha)$).

Numerical evidence of the so-obtained results are reported in section \ref{sec_num_ex}.

\section{Order statistics, tail exponent estimation and multifractal formalism}
\label{sec_tail}
\subsection{Notations}
The idea underlying tail exponent estimation 
is to study statistics of the $k(N)$ observed extreme values 
in the asymptotic regime $N \rightarrow +\infty$ 
and $k(N) \rightarrow +\infty$. As in the previous section, the limit $N \rightarrow +\infty$ is taken 
as explained in section
\ref{sec_mix}, using the $\chi$ exponent (see Eqs (\ref{Ntau}) and (\ref{defchi})).

For the $k(N) \rightarrow +\infty$ limit, it is convenient to parametrize $k(N)$ as:
\begin{equation}
\label{defnu}
  k(N) \sim  N^{\nu}
\end{equation}
where $0 \leq \nu \leq 1$, the value $\nu = 0$ being interpreted as
$k(N) \sim \ln(N)$. We will denote
$\hat{\mu}(\nu,\chi)$ the estimated tail exponent using
one of the estimators reviewed in section \ref{sec_ev} for
some  given $\chi$ and $\nu$ ($\chi$ is defined in the section \ref{sec_mix}).
At scale $\tau = T2^{-n}$, as in section \ref{sec_ev}, we denote
$X_1 \ldots X_N$ the rank ordered values of $M(I_n)$ over dyadic intervals.

\subsection{Tail exponent estimators}
Let us compute the expected value of the 
tail exponent estimator (Power-law fit, Pickands or Hill estimator) 
for fixed values of $\nu$
and $\chi$. For the sake of simplicity we focus on Pickands estimator
but the same argument equally applies to other estimators (e.g. Hill).
Our heuristics will rely upon the multifractal formalism.

 As recalled in section
\ref{mf},  the positive part of 
the Legendre transform of $\zeta(q)$ corresponds 
to the singularity spectrum, i.e., the Haussdorf dimension
of sets of iso-regularity. Mandelbrot has proposed a long time ago,
 a probabilistic interpretation of {\em negative dimensions} (i.e., negative values of $f(\alpha)$)
in terms of large deviation spectrum \cite{Man90}.
In this section, we will use this formalism in order to 
study the order statistics of multifractal fluctuations.

The Pickands estimator is simply defined as:
\begin{equation}
\label{pick}
  \hat{\mu}(\nu,\chi) = \ln(2) \left(\ln \frac{X_k-X_{2k}}{X_{2k}-X_{4k}}
  \right)^{-1} 
\end{equation}

Let us define $\alpha_{\nu,\chi}$ such that
\[
  k = N^{\nu} \sim N_T \left(\frac{\tau}{T}\right)^{-f(\alpha_{\nu,\chi})}
\]
Using the definition of $\chi$ according to which $N_T \sim (\tau/T)^{-\chi}$,
it follows that $\alpha_{\nu,\chi}$ satisfies
\begin{equation}
\label{defhnu}
  f(\alpha_{\nu,\chi}) = \nu-\chi(1-\nu)
\end{equation}

Let us now consider $\alpha'_{\nu,\chi} = \alpha_{\nu,\chi}+\epsilon_1$ such that
\[
  2 k \sim N_T \left(\frac{\tau}{T}\right)^{-f(\alpha'_{\nu,\chi})}
\]
i.e., $2 k \sim  N_T \left(\frac{\tau}{T}\right)^{-f(\alpha_{\nu,\chi})+\epsilon_1 q_{\nu,\chi}}$, thus $2^{-1/q_{\nu,\chi}} \sim \left(\frac{\tau}{T}\right)^{\epsilon_1}$
where 
\begin{equation}
\label{qnu}
q_{\nu,\chi} = \frac{d f}{d \alpha}(\alpha_{\nu,\chi}) \; .
\end{equation}
Along the same line if 
$\alpha''_{\nu,\chi} = \alpha_{\nu,\chi}+\epsilon_2$ such that
\[
    4 k \sim N_T \left(\frac{\tau}{T}\right)^{-f(\alpha''_{\nu,\chi})}
\]
we have $4^{-1/q_{\nu,\chi}}  \sim \left(\frac{\tau}{T}\right)^{\epsilon_2}$.
Thanks to the fact that, 
\begin{eqnarray*} 
  X_k -X_{2k} & \sim & X_k \left(1-(\tau/T)^{\epsilon_1}\right) \\ 
  X_{2k} -X_{4k} &  \sim & X_k
  \left((\tau/T)^{\epsilon_1}-(\tau/T)^{\epsilon_2}\right)
\end{eqnarray*}
one finally gets for expression (\ref{pick}):
\begin{equation}
\label{munu}
\hat{\mu}(\nu,\chi)  \simeq \ln(2) \left(\ln 
\frac{1-2^{-1/q_{\nu,\chi}}}{2^{-1/q_{\nu,\chi}}-4^{-1/q_{\nu,\chi}}} \right)^{-1} = q_{\nu,\chi}
\end{equation}
We then come to the conclusion that the Pickands tail estimator for a
multifractal
process strongly depends on the choice of the rank $k=N^{\nu}$ and 
the exponent $\chi$.
This is another difference with standard theory for i.i.d. random
variables. The same kind of phenomenology can also be applied to the
Power-law fit or Hill estimator.
Let us notice that when $\nu = 0$ we consider only the 'extreme' of
the 'extremes' for the tail exponent estimation and therefore 
we recover the tail of the law of the maximum value as discussed previously.
Indeed, from Eqs. (\ref{defhnu}), (\ref{qnu}) and (\ref{munu}), we get
\begin{eqnarray*}
   \hat{\mu}(0,\chi) & = & f'(\alpha_{0,\chi}) \\
   f(\alpha_{0,\chi}) & = & -\chi \; ,
\end{eqnarray*}
that is, exactly Eqs. (\ref{tailmax}).

Numerical evidence illustrating these results are provided in the next section.

\section{Numerical examples}
\label{sec_num}

Most of our considerations rely on phenomelogical scaling
and asymptotic limit argmuments. We therefore neglected prefactors and 
slowly varying corrections, i.e., finite size effects.
In order to test our approach as well as to quantify the 
importance of finite-size corrections, it is therefore
interesting to perform numerical simulations.
In this section, we illustrate our purpose  
on specific examples such as those mentionned in 
Appendix \ref{ap_excasc}. As already explained, 
most of these simulations are performed on continuous cascades which
construction algorithm is described in ref. \cite{MuBa02}.

\begin{figure}
\hskip -0.5cm \includegraphics[height=8.5cm]{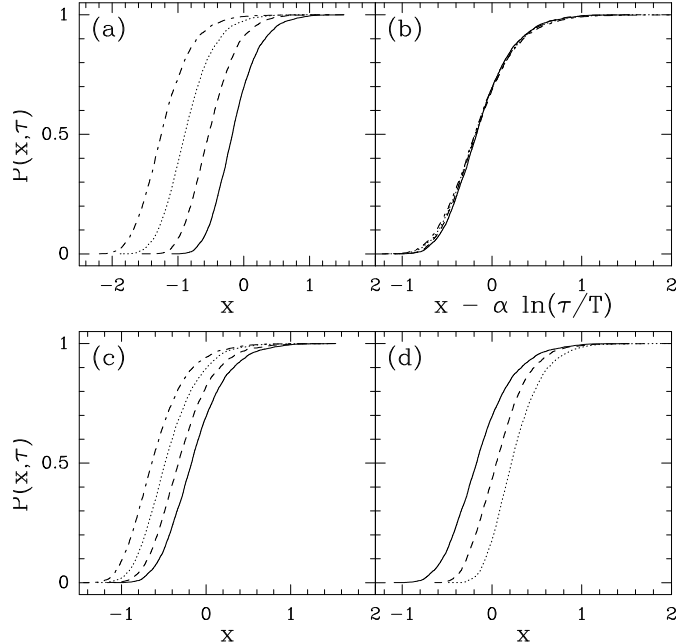}
\caption{\label{figA} Cumulative probability distribution
of the maximum of $\ln(M)$ at scale $\tau = T2^{-n}$, $P(x,n)$ 
(also referred to as $P(x,\tau)$), as a function of $x$ for
various values of $\tau$ for a continuous log-normal
cascade with $\lambda^2 = 0.2$. (a) $N_T = 1$ and $\ln_2(T/\tau) = 3$ (continuous line), 4, 5, 6
(dotted lines). According to the formalism of section \ref{sec_mfev}, 
the cdf should be a front moving toward $x < 0$ at a
``velocity'' $\alpha_{\star,\chi=0} = \alpha_{\star} \simeq 0.47$. (b) All the cdf of Fig. (a) merge to a single
curve when plotted in the ``moving referential''. (c) Same plot as Fig. (a) but with $\chi = 1$.
One expects a smaller velocity $\alpha_{\star,\chi=1} \simeq 0.2$. (d) Same plot as figs. (a) and (c) for scales
$\ln_2(T/\tau) = 3,4,5$ with $\chi = 3$. One observes, as expected, a negative velocity $\alpha_{\star,\chi=3} \simeq  -0.16$.} \end{figure}

\begin{figure}
\hskip -0.5cm \includegraphics[height=5.5cm]{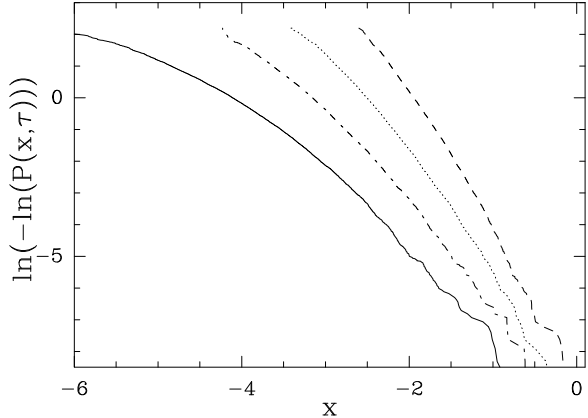}
\caption{\label{figB} ``Gumbel plots'' of the cumulative probability distribution
 of the maximum of $\ln(M)$ at scale $\tau = T2^{-n}$, $P(x,n)$ (also referred to as $P(x,\tau)$) for different values of $\chi$. 
One sees that as $\chi$ increases from $0$ to $1.5$, $q_{\star,\chi}$, the slope in the tail
increases significantly.} 
\end{figure}

\subsection{Extreme value statistics}
\label{sec_num_ex}
In order to check the results of section \ref{sec_mfev}, we have
generated $N = 5000$ independent realisations of log-normal continuous
cascades of intermittency parameter $\lambda^2 = 0.2$ for various values of 
the parameters $N_\tau$ and $N_T$. From these independent samples, we have 
estimated the cdf of the maximum value of $\ln(M)$ at scale $\tau = T2^{-n}$.
In Fig. \ref{figA}(a) the cdf $P(x,n)$ (also referred to as $P(x,\tau)$) is plotted for $N_T = 1$ 
and $T/\tau = 8,16,32,64$. According to Eqs. (\ref{gumbelfront}) and (\ref{tailmax}), we expect
to observe, as the scale $\ln(\tau/T)$ decreases, a front traveling towards negative $x$ at velocity
$\alpha_{\star,\chi=0}=\alpha_{\star}$. This behavior is well verified and
can be quantitatively checked in Fig. \ref{figA}(b) where all the fronts merge 
when plotted versus
$x -\alpha_{\star} \ln(\tau/T)$ with $\alpha_{\star} \simeq 0.47$ as given according to Eqs. (\ref{tailmax}) and (\ref{ln}):
\[
  \alpha_{\star} = 1+\frac{\lambda^2}{2}-\lambda \sqrt{2(1+\chi)} \; .
\]
In Fig. \ref{figA}(c) the same analysis is performed for $N_T = T/\tau$ and
therefore $\chi = 1$. One observes that the velocity decreases as expected
(in that case $\alpha_{\star,\chi=1} \simeq 0.2$). For $\chi$ large enough, $\alpha_{\star,\chi}$ can
become negative as illustrated in Fig. \ref{figA}(c) where we have reproduce the plot of Fig. \ref{figA}(a) for
$\tau = 8,16,32$ and $\chi = 3$. According to previous formula, $\alpha_{\star,\chi=3} \simeq -0.16$, a value compatible
with observations where one sees the front moving towards positive $x$ values as the scale decreases. 
The fact that $\alpha_{\star,\chi}$ is negative for $\chi$ large enough can be easily understood as follows:
because the measure is continuous, as $\tau$ decreases, the maximum is expected to go to zero, but
if in the same time, the number of indenpendent integral scales is increased, the maximum is expected to increase. The
exponent $\chi$ controls the balance between these two opposite effects. For $\chi$ large enough one explores 
negative dimensions that can be associated with negative $\alpha$'s.
Notice that in all plots one observes a slight change 
in the shape of the front as the scale goes to zero: this 
is not surprising because the asymptotic shape of the front depends on $\chi$ and is a priori different
from the initial front at scale $T/\tau = 8$ considered in Fig. \ref{figA}. 

According to Eq. (\ref{gumbelfront}), the asymptotic shape of the 
fronts should be Gumbel, i.e.,
\[
  P(x) = e^{-Ce^{-q_{\star,\chi}x}}
\]
 with parameter $q_{\star,\chi}$ for the log-normal model:
\[
 q_{\star,\chi} = \sqrt{{\frac{2(1+\chi)}{\lambda^2}}} \; .
\]
Notice that $q_{\star,\chi}$ is an increasing function of $\chi$.
In Fig. \ref{figB}, we have plotted $\ln(-\ln(P(x))$ versus $x$ for various values of 
$T/\tau = N_\tau$ and $N_T$. We have chosen the values $(N_\tau = 512, N_T = 1)$, $(N_\tau = 512, N_T = 8)$, $(N_\tau =512,N_T = 64)$ and
($N_\tau = 64, N_T = 512)$ that correspond respectively (if one neglects prefactors in our analysis) to $\chi = 0$, $\chi = 0.33$, 
$\chi = 0.67$, $\chi = 1.5$.
For an exact Gumbel law, one would expect a straight line. One clearly observes strong deviations to the Gumbel shape
because the asymptotic regime is not reached but the tails look straight and are very well estimated by the theoretical
$q_{\star,\chi}$ values associated with the values of $\chi$: one sees a systematic increase of $q_{\star,\chi}$ when one goes from
$\chi= 0$ to $\chi = 1.5$. 
 
\begin{figure}
\hskip -0.7cm \includegraphics[height=8.5cm]{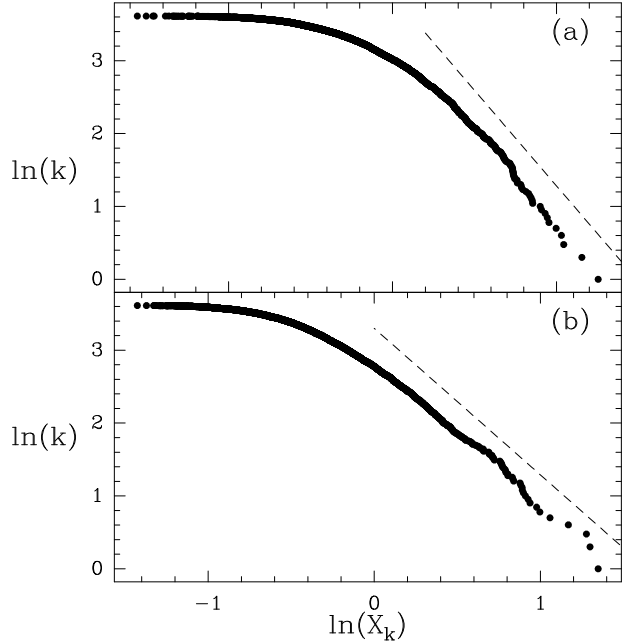}
\caption{\label{fig4} Rank-Frequency (Zipf) plots in log-log representation
for a log-Normal cascade (top) and a log-Gamma cascade (bottom). The slope of the 
right linear part provides an estimate of the tail exponent $\mu$. Dashed lines
indicate analytical expectations (see text).}
\end{figure}

\subsection{Tail exponent estimators}
\label{sec_num_tail}
Let us now check our results on the tail behavior of the estimated pdf
from measure samples.

\begin{figure}[t]
\hskip -1cm \includegraphics[height=8.5cm]{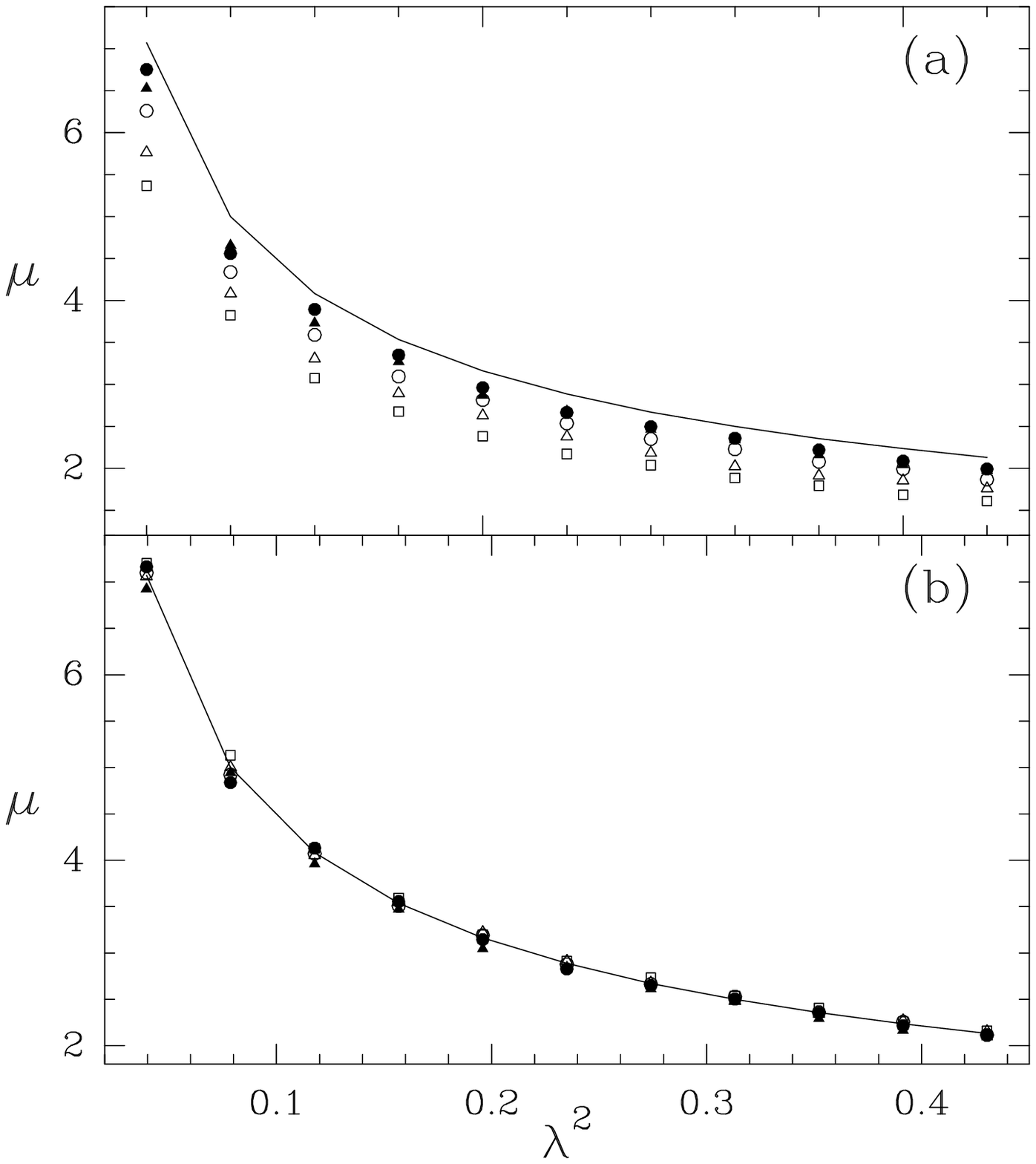}
\caption{\label{fig5} Mean estimated tail exponent as a function
of the intermittency coefficient $\lambda^2$ 
for continuous log-normal cascades.
Each mean value is computed using 1000 cascade samples of length $L$.
(a) Power-law fit estimator as a function of $\lambda^2$ for
$T = 1024$, $L = 4096$, $k = 16$ ($\bullet$), $k= 32$ ($\circ$),
$k = 64$ ($\triangle$), $k= 128$ ($\Box$) and $T=512$, $L=8192$,
$k=32$ ($\blacktriangle$). The continuous line corresponds to the theoretical
prediction $\mu_{ln} = \sqrt{2/\lambda^2}$ (Eq. (\ref{muln})).
(b) The same as in (a) but each curve has been rescaled by
a factor $\sqrt{2(1-\nu)(1+\chi)}$ according to Eq. (\ref{muln}) (see text).}
\end{figure}

\begin{figure}
\hskip -1cm \includegraphics[height=4.5cm]{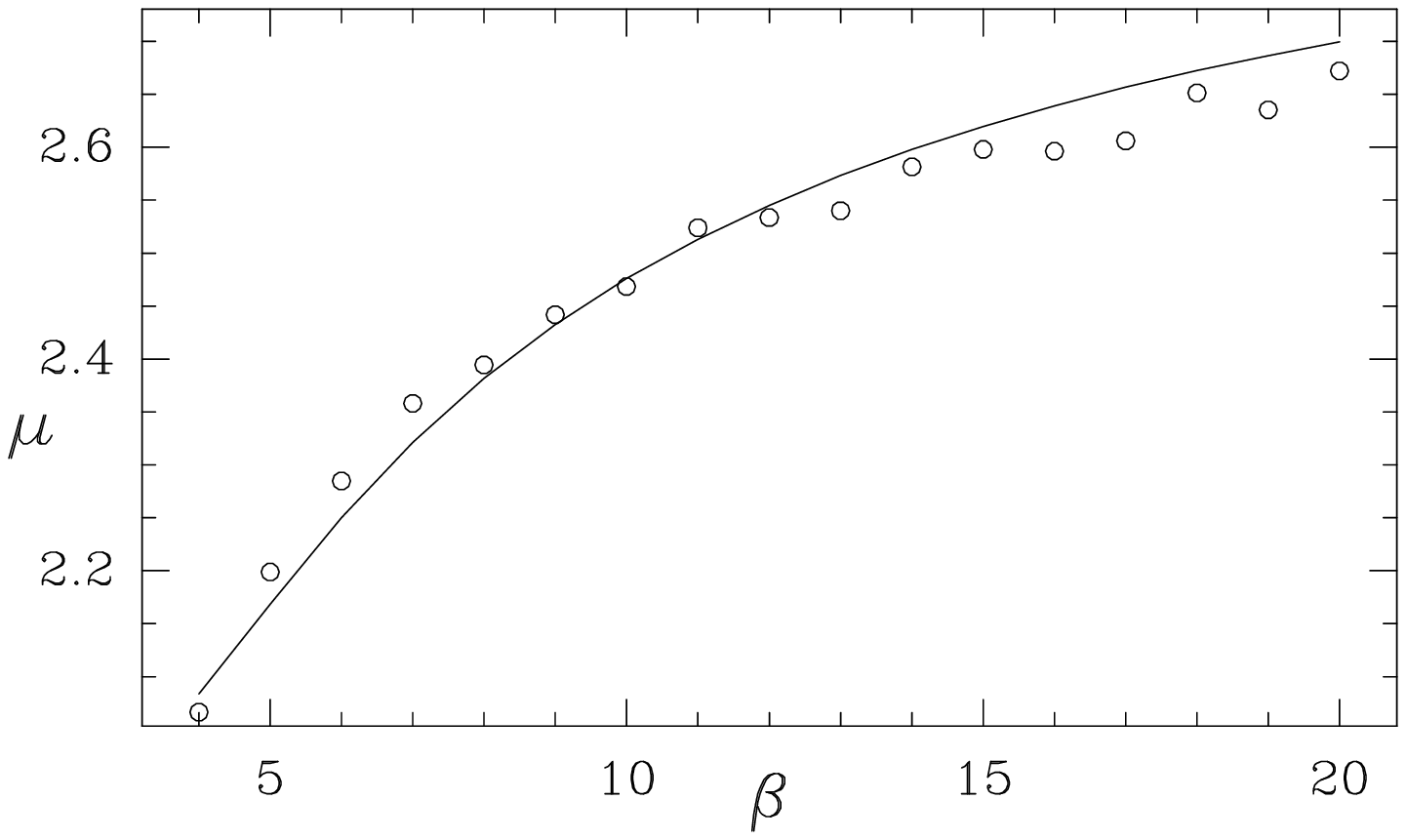}
\caption{\label{fig6} Mean estimated (Power-law fit) tail exponent ($\circ$) 
as a function of $\beta$ for discrete log-Gamma cascades with $\lambda^2 = 0.2$ and
$T = 2048$. The Hill parameter is $k = 32$ and each mean value is
computed from 1000 cascade samples. The continuous line corresponds
to the analytical expression (\ref{mulg})}.
\end{figure}

Let us first compute $\hat{\mu}_{\nu,\chi}$ as given by Eq. (\ref{munu}), i.e. solve
Eqs. (\ref{defhnu}) and (\ref{qnu}) 
for respectively log-Normal, log-Poisson and log-Gamma
examples. After some simple algebra, we find in the log-normal case
\begin{equation}
\label{muln}
  \hat{\mu}_{ln}(\nu,\chi)  = \sqrt{\frac{2(1+\chi)(1-\nu)}{\lambda^2}}
\end{equation}
Let us notice that for $\nu = 0$, one recovers previous tail exponent
of the extreme values $\hat{\mu}(0,\chi) = q_{\star,\chi}$.
In the case of log-Poisson statistics (Eq. (\ref{zetalp})), 
the computation leads to an expression for $\hat{\mu}(\nu,\chi)$
that involves Lambert $W(x)$ function.
A pertubation series in the
limit $\delta \rightarrow 0$ of this expression gives:
\begin{equation}
 \label{mulp}
  \hat{\mu}_{lp}(\nu,\chi) = \sqrt{\frac{2(1+\chi)(1-\nu)}{\lambda^2}}+2 \delta \frac{(1+\chi)(\nu-1)-1}{3 \lambda^2} + \ldots
\end{equation}
Along the same line, the value of $\hat{\mu}(\nu,\chi)$ can
be computed in the case of the log-gamma cascade (Eq. (\ref{zetalg})) which
again involves the second branch of the Lambert function $W_{-1}$:
\begin{equation}
\label{mulg}
  \hat{\mu}_{lg}(\nu,\chi)  =
  \beta\left[ 1-e^{1+W_{-1}\left(-e^{-1+\frac{(1+\chi)(\nu-1)}{\lambda^2\beta^2}}\right)
  +\frac{(1+\chi)(1-\nu)}{\lambda^2\beta^2}} \right] 
\end{equation}
A series expansion in the limit $\beta \rightarrow +\infty$ gives:
\begin{equation}
\label{muld}
  \hat{\mu}_{lg}(\nu,\chi) = \sqrt{\frac{2(1+\chi)(1-\nu)}{\lambda^2}}
+4 \frac{(\nu-1)(1+\chi)}{3 \beta \lambda^2} + \ldots
\end{equation}

In Fig. \ref{fig4}, the rank ordering
of the a log-normal and a log-Gamma (with $\beta = 4$) cascade processes with
an intermittency coefficient $\lambda^2= 0.2$ 
are plotted in doubly logarithmic scale (such
plots are often referred to as ``rank-frequency'' plots or ``Zipf''
plots \cite{BSor,New04}). 
One clearly sees that the rightmost part of each distribution behaves
as a power law. From Eqs. (\ref{muln}) and (\ref{mulg}) (with $\chi = 0$ and $\nu = 0$),  
the slope of the plots should be respectively $\mu \simeq 3.1$ and
$\mu \simeq 2.0$. One sees that these behaviors, reported 
on the figures as dashed lines, fit the data relatively well. 
One can observe that the the scaling range associated with the 
log-Gamma cascade is wider than for the log-Normal measure. Indeed,
according to Appendix \ref{sec-last} (see Eq. (\ref{srange})), in the log-Normal case, this
range should be around $0.2 pq_\star \ln(T/\tau) \simeq 0.7$ for $p = 0.1$ while
its values for log-Gamma is expected to be $1.13 pq_\star \ln(T/\tau) \simeq 2$ for $p = 0.1$, i.e.,
more than two times wider than for the log-Normal case. This difference can be visually 
checked in Fig. \ref{fig4}.

In Fig. \ref{fig5}, we have plotted the mean value of the
tail exponent estimator (power-law fit estimator) as a
function of the intermittency parameter $\lambda^2$ in 
the case of a log-Normal (continuous) cascade for various
values of $k$, $N_T$ and $N_\tau$. The theoretical prediction (\ref{muln})
for $\nu = 0$ is shown in continuous line for comparison.
The mean has been evaluated on 
$10^3$ independent cascade samples. In Fig. \ref{fig5}(a), one
can see that for all parameter values, the curves have the same
decreasing behavior. All these curves collapse on the theoretical
prediction $\sqrt{2/\lambda^2}$ as shown in Fig. \ref{fig5}(b), 
if one rescales each one by a factor $\sqrt{(1-\nu)(1+\chi)}$ 
where $\nu$ and $\chi$ are computed from expressions (\ref{defnu}) and (\ref{defchi})
by assuming that the prefactors are trivially 1.
Empirically we find that the asymptotic phenomenology works quite well and
prefactors or slowly varying behavior are negligible.
As previously emphazised, the
exact computation of prefactor values and finite size effects is beyond
the scope of this paper and should involve more sophisticated
mathematical tools (see appendix \ref{ap_results}).

In Fig. \ref{fig6}, we compare the expression (\ref{mulg}) to 
the behavior of the tail
exponent estimator as a function of $\beta$ in the 
case of a discrete log-Gamma cascade: in that case, $\lambda^2$ is fixed ($\lambda^2 = 0.2$)
while $\beta$ varies. Once again, we can see that
if we take into account finite $k$ value through 
the value of $\nu$ and $\chi$ close their expected values, the analytical
expression provides a very good fit of the data. 

We have therefore illustrated, on two specific examples, that
the phenomenology developed in previous section allows us
to predict with a relative precision 
the tail behavior of multifractal measure samples.
Let us now show how financial time series fluctuations
can be described within this framework.

\section{Application to financial data}
\label{sec_finance}

Multifractal models are, with many regards, 
well suited to account for return fluctuation of
financial assets \cite{Bman,BBoPo}. Among the
``stylized facts'' characterizing the asset return fluctuations,
the phenomenon of ``volatility clustering'' (called heteroskedaticity
in econometrics) is the most important one. One of the
key points raised in refs. \cite{MuDeBa00,BaDeMu01} is
that these volatility correlations are found empirically very close
to the ``log-correlations'' of continuous multifractal cascades \cite{AMS98}.
Therefore the observed multiscaling properties of returns
can be simply explained in terms of  volatility persistence.
In this section we want to stress that, for the same reason, 
namely the logaritmic shape of log-volatility correlations, the pdf
of return fluctuations appear empirically as fat tailed with
a rather small tail exponent. 

In ref. \cite{MuDeBa00,BaDeMu01}, we have shown
that a parcimonious model of $X(t)$, some asset return value at time 
$t$, can be 
constructed as follows:
\begin{equation}
\label{subord}
  X(t) = B(M(t))
\end{equation}
where $B(t)$ is the standard Brownian motion and $M(t)$ is 
a multifractal continuous cascade as defined in refs. \cite{MuBa02,BaMu03}:
\[
  M(t) = \lim_{l \rightarrow 0} \int_0^t e^{\omega_l(t)} dt \; .
\] 
The process $\omega_l(t)$ plays exactly the same role as $\omega = \ln(W)$
in discrete cascades.
It is easy to see that the variations of $M(t)$ in (\ref{subord}) 
can be interpreted as a stochastic variance. This quantilty is called 
the ``volatility'' in finance \cite{BBoPo}.

If $M(t)$ has multiscaling (e.g. log-normal) properties then so do $X(t)$. 
Empirically, it has been determined using data from 
several markets, from various countries, that the intermittency coefficient
charaterizing the multifractal statistics of the volatility is
close to $\lambda^2 = 0.2$ while the integral scale $T$ is typically 
around $1$  year \cite{MuDeBa00}.
On the other hand, 
many studies relying on high frequency data or on thousands of
daily stock returns, have revealed that the financial return pdf
have heavy tails with a tail exponent $\mu$ in the interval $[3,5]$. 
This is the famous ``inverse cubic'' 
law for return fluctuations \cite{stanley,BBoPo}.
This observation lead to one of the main objections 
raised against the previous multifractal
model for asset returns \cite{BBoPo}. 
Indeed, the {\em unconditional pdf} 
of the volatility associated with a log-normal multifractal
cascades of coefficient $\lambda^2 = 0.15$ has
a tail exponent $\mu \approx 13$ (Eq. (\ref{muln1})).
Within the ``subordinated'' model (\ref{subord}), this would mean that
the tail of the volatity pdf is around $2 \mu \approx 26$, i.e.,
close to ten times the observed value !
However, the main message of this paper is that for multifractal 
fluctuations, the observed extreme events are far from being
distributed as they were independent. In particular we have
shown in section \ref{sec_tail} that the estimators of the pdf tail
exponent strongly depend on $\nu$ and $\chi$ and are ``generically''
smaller than $\mu$. Typically, in finance
$\tau \simeq 10^{-2}-1$ day, $T \simeq 1-2$ years and $L \simeq 10$ years.
Therefore, a rough estimate of $\chi$ and $\nu$ can be
$\chi \simeq \nu \simeq 0.5$.
If one uses these values in Eq. (\ref{muln}), one finds a typical 
value for the estimated tail within the log-normal model that is 
$2 \hat{\mu} \simeq 6$.
This is a value closer to the observations. In order to have 
a better fit of the tail behavior one could use a log-gamma
model with the same intermittency coefficient and $\beta = 4$
(see Appendix \ref{ap_excasc}). In that case, Eq. (\ref{mulg})
gives an estimator value $2 \hat{\mu} \simeq 3.6$ that agrees
with observations.

\begin{figure}
\hskip -0.7cm \includegraphics[height=4.5cm]{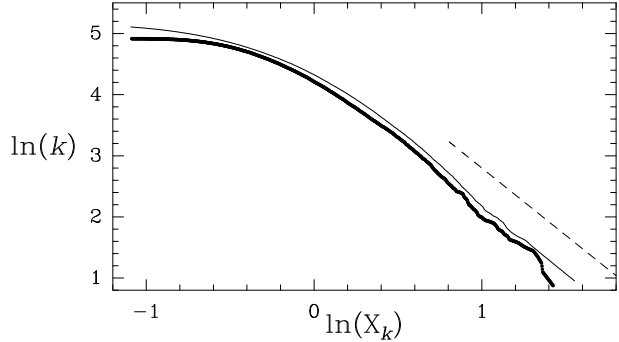}
\caption{\label{fig8} Rank-Frequency plot of CAC40 daily volatility
estimates ($\bullet$) as compared to simililar plot for
a log-Gamma continuous cascade with $T = 253$ days, $\lambda^2 = 0.05$
and $\beta = 4$ (thin line). The fit of the extreme tail provides an estimation
$\mu \simeq 2$ (dashed line).}
\end{figure}

In Fig. \ref{fig8} is reported a rank-frequency plot of
estimated daily volatilities associated with the 40 stock values 
composing to the French CAC 40 index. The data are daily 'open' 'high'
'low' 'close' quotes over a mean period of 10 years. The daily
volatilities are estimated using the widely used 
Garman-Klass method \cite{GK} and each volatility 
sample mean has been normalized to 1.
For comparison the Zipf plot associated with a multifractal
log-Gamma measure with $T= 1$ year, $\lambda^2 = 0.2$ and $\beta = 4$
has been also reported. One can see that both curves behave very
similarily with a power-law tail exponent $\mu \simeq 2$. Let us note
that for small volatilites values, the behavior of CAC40 volatility
pdf is slightly different from the log-Gamma cascade probably because
of the hight frequency noise in the Garman-Klass volatility estimates.
This figure illustrates very well our result: there is no discrepency 
between the value of the intermittency
coefficient and the estimated pdf power-law behavior.

\section{Summary and prospects}
\label{sec_conclusion}
In this paper we have addressed the problem of extreme value statistics
for multifractal processes.
This problem is non trivial and possesses
a rich phenomenology involving non ergodic
behavior. In the case of multifractal
processes, two important parameters govern the asymptotics: the 
overall sample lenght and the scale at which data are sampled.
The exponent $\chi$ we have introduced, precisely quantifies how one
defines the asymptotic limit as one changes these two parameters.
The observed extreme value statistics result from a ``competition'' between
an increase of the number of independent samples (which tends to 
increase typical extreme values) and a decrease of the sampling scale
(which tends to decrease typical observed values). Consequently,
the law of extremes continuously depends on $\chi$, 
an exponent that turns out to be interpreted as a {\em negative dimension}.

This exponent naturally plays an important role when one wants
to estimate the tail of the probability 
law associated with multifractal fluctuations.
Using the phenomenology of the multifractal formalism we have shown 
that tail exponent estimators continuously depend on $\chi$. They also 
depend on another exponent $\nu$ that quantifies ``how many'' extremes
one uses for estimation. Under usual experimental
conditions whe have notably shown that the obtained exponent is usually 
smaller than the exponent expected from the unconditional law.
Such non ergodic behavior are similar to those observed
in the thermodynamics of disordered systems, at low temperature, under
the freezing transition \cite{DerSpo88,CarLeDou01}.

The argmuments and methods we used in this paper are mostly phenomenological
and rely upon large deviation type arguments, multifractal formalism
and traveling front solution of non linear iteration equations.
Beyond the need for sitting it on rigorous bases, they
are other appealing mathematical prospects raised by our appoach, such
as the possibility to address finite size effects in multifractal
scaling laws or to define a precise statistical framework for
studying them.

As far as applications are concerned, we have provided
a direct use of our results in the field of econophysics
where multifractal models for asset returns are popular.
Whe have shown that the observed fat tails of return
pdf are well reproduced by a multifractal model designed
to account for the volatility clustering phenomenon.
Other fields where multifractal processes are involved
can be potentially investigated along the same line.
Conversely and perhaps more importanlty, 
within this framework, multifractality  
appears as an alternative that 
can invoked to explain the origin of fat tails as observed 
in many fields of applied science \cite{BSor,BouPL01,New04}.

\begin{acknowledgments}
We thank J.P. Bouchaud for helpful discussions
about the analogy of some multifractal problems addressed in this paper and 
similar problems arising in physics of disordered systems.
This work as been supported by contract ACI {\em nouvelles
interfaces des math\'ematiques}, Nb 54-03, from french minist\`ere
de l'\'education nationale. 
\end{acknowledgments}

\appendix
\section{Log-normal, log-Poisson and log-Gamma multifractal measures}
\label{ap_excasc}
In this appendix we provide 3 examples of 
multifractal statistics to which we will refer all along the paper. 
We consider the 3 infinitely divisible 
laws of $\ln W$ (or equivalently $\omega$): normal,
Gamma and Poisson.

In the simplest case, $\omega = \ln W$ is normal 
of variance $\lambda^2/\ln(2)$
(and mean $-\lambda^2 /2\ln(2)$ because $E(W) = 1$). Then
$\EE{W^q} = e^{-q \lambda^2/2 +\lambda^2 q^2/2}$ and we recover 
the expression of Eq.~(\ref{ln}):
\begin{equation}
\label{zetaln}
  \zeta_{\mbox{ln}} (q) = q(1+ \lambda^2/2) - \lambda^2 q^2/2
\end{equation}
The intermittency coefficient in this case is simply $\lambda^2$. If one
solves $\zeta(q) = 1$, $q>1$ one gets 
\begin{equation}
\label{muln1}
\mu = \frac{2}{\lambda^2}.
\end{equation}
In the second example, $n$ is a Poisson random variable of intensity $\gamma \ln(2)$ and 
$\omega = m_0 \ln(2) + n \delta$. Then 
$q-\ln_2 \EE{e^{q \omega}}= q(1-m_0)+\gamma(1-e^{q\delta})$. If one 
sets $\zeta(1) = 1$ and $-\zeta''(0) = \lambda^2$, one obtains the 
spectrum: 
\begin{equation}
\label{zetalp}
  \zeta_{\mbox{lp}} (q) = q\left(1+ \frac{\lambda^2}{\delta^2}(e^\delta-1)\right) + \frac{\lambda^2}{\delta^2}\left(1-e^{q\delta}\right)
\end{equation}
Notice that when $\delta \rightarrow 0$, $\zeta_{\mbox{lp}} \rightarrow \zeta_{\mbox{ln}}$. It is easy to show that, when $\delta$ is negative and small enough, 
for all $q>1$, $\zeta_{\mbox{lp}}(q) > 1$
and therefore $\mu = +\infty$ in that case.
 
In the third example, $\omega$ is Gamma distributed: if $x$ is a random 
variable of pdf $\beta^{\alpha \ln(2)} x^{\alpha \ln(2) -1} e^{-\beta x} /\Gamma(\alpha \ln(2))$ and
$\omega = x+m_0 \ln(2)$, then $q-\ln_2 \EE{e^{q \omega}} = q(1-m_0)+\alpha \ln\left(1-q/\beta\right)$. By setting $\zeta(1) = 1$ and $-\zeta''(0) = \lambda^2$,
we have:
\begin{equation}
\label{zetalg}
 \zeta_{\mbox{lg}}(q) = q \left(1-\lambda^2\beta^2\ln(\frac{\beta-1}{\beta})\right)
+ \lambda^2 \beta^2 \ln\left(\frac{\beta-q}{\beta}\right)
\end{equation}

Notice that $\zeta_{\mbox{lg}} \rightarrow \zeta_{\mbox{ln}}$ when $\beta \rightarrow +\infty$. The solutions of $\zeta_{\mbox{lg}}(q) = 1$ can be obtained in
terms of Lambert $W$ function (satisfying $W(x)e^{W(x)} = x$) and therefore
the value of $\mu$ can be exactly computed as a function of $\lambda^2$ 
and $\beta$.

\section{Proof of Eq.\ref{iter}}
\label{ap_iter}
In this appendix we prove Eq. (\ref{iter}). 
Let us first study how the law of the maximum of $\ln(M(I_{n})/\tau)$ 
inside one integral scale varies when on changes the scale $\tau = T 2^{-n+1}$ 
to $\tau = T 2^{-n}$.
Let $I_n(k)$, $k = 0 \ldots 2^n-1$
denote the dyadic intervals of size $T 2^{-n}$ of the interval $[0,T]$.
From the cascade construction, $\forall n$, 
the following stochastic equality can be easily proven:
\begin{eqnarray*}
    & M\left[I_{n}(k)\right]  &  \operatornamewithlimits{=}_{fdd}   \frac{W_1}{2} M_1(I_{n-1}(k)) \;,\; k \in  [0,2^{n-1})  \\ 
   & M\left[I_{n}(k)\right]  & \operatornamewithlimits{=}_{fdd} \frac{W_2}{2} M_2(I_{n-1}(k-2^{n-1})) \;,\;  k \in [2^{n-1},2^{n}] 
\end{eqnarray*}
where $W_1$, $W_2$, $M_1$ and $M_2$ are independent copies
of $W$ and $M$ respectively. The symbol $fdd$ means an equality in law for
all finite dimensional distributions.
Therefore, if 
\[
 X(n) = \max_k \left[\ln(M(I_n(k))/\tau)\right] 
\] 
One has 
\[
 X(n) \operatornamewithlimits{=}_{law} \max \left[
 \ln(W_1)+X_1(n-1),\ln(W_2)+X_2(n-1)\right]
\]
If $g(x)$ denote the law of $\omega = \ln(W)$ and $P'(x,n) = \PP{X(n) > x}$,
the previous equality can be rewritten as
\begin{eqnarray}
\nonumber
  P'(x,n) &  = &\left[ \int P'(z-x,n-1) g(z) \; dz \right]^2 \\
\nonumber
  &  = & \left[ g \star P'(x,n-1) \right ]^2
\end{eqnarray}  
Now if one has $N_T = 2^{rm} > 1$ integral scales , the cdf of the 
maximum is given by (\ref{iter}).

\section{Aronson-Weinberger criterium and finite size effects on front solutions of Eq. (\ref{iter})}
\label{ap_results}

In this Appendix we provide some additional technical details on
the solutions of Eq. ({\ref{iter}). We do not establish rigorous proofs but
mostly recast some results from refs. \cite{VS03,Brunet,BruDer97,MajKra00,DeMaj01,MajKra02} to our problem.
If one linearizes Eq. (\ref{iter}) in the tail region $x \rightarrow +\infty$, one
obtains the following recursion for $Q(x) = 1-P(x,m)$: 
\begin{equation}
\label{linearT}
  Q(x,m+1) = 2^{p+r} Q(x,m) \ast g^{(p)}(x) 
\end{equation}
where $g^{(p)}(x)$ is simply the product of $p$ convolutions $g(x) \ast \ldots \ast g(x)$.
If one decomposes $Q(x,m)$ on Fourier modes:
\[
  \hat{Q}(k,m) = \int e^{-ikx} Q(x,m) \; dx
\]
then (\ref{linearT}) becomes:
\[
  \hat{Q}(k,m+1) = 2^{p+r} \hat{Q}(k,m) e^{p F(ik)}
\] 
where $F(ik) = \ln \EE{e^{-ik\omega}}$ is the cumulant generating function of $\omega = \ln(W)$ the
logarithm of cascade weights. The solution is therefore
\[ 
  \hat{Q}(k,m) = A(k) e^{m\ln(2)\left[r+p(1+F(ik)/\ln(2))\right]} 
\]
where $A(k)$ is simply the Fourier transform of the initial condition.
$Q(x,m)$ is obtained as the inverse Fourier transform:
\[
  Q(x,m) = (2 \pi)^{-1} \int e^{ikx} A(k) e^{m\ln(2)\left[r+p(1+F(ik)/\ln(2))\right]} \; dk
\]

In a referential moving at velocity $v$, (i.e. $x_m = x_0+vm$), the previous integral 
can be computed using a steepest descent method: One deforms the integral over the real axis 
to a contour in the complex plane of constant phase. 
In the  limit $m \rightarrow +\infty$, the main contribution comes from saddle point of the 
function $F(ik)$ along this path:
\[
  Q(x_m,m) \sim A(k_{\star}) e^{ik_{\star} x_m + m\ln(2)\left[r+p(1+F(ik_{\star})/\ln(2))\right]}
\]
where $k_{\star}$ satisfies: 
\begin{equation}
\label{statphase}
  -i k_{\star} v = \frac{d F(ik)}{d k}|_{k=k_\star}
\end{equation}
Moreover, if the front is stationary in the moving frame, the selected velocity 
should be such that the real part of the exponent is zero:
\begin{equation}
\label{rezero}
\Re \left\{ik_{\star} m v + m\ln(2)\left[r+p(1+\frac{F(ik_{\star})}{\ln(2)})\right]\right\} = 0
\end{equation}
and thus, if $v$ is real, by setting $k = iq$, 
and thanks to the equality
\[
  F(-q) = \ln(2)(q-\zeta(q))
\]
one can rewrite respectively Eqs. (\ref{rezero}) and (\ref{statphase}) as follows: 
\begin{eqnarray*}
  & v(q)  =  \frac{\ln(2)[r+p(1+q-\zeta(q))]}{q} \\
  & \frac{\partial v(q)}{\partial q}|_{q=q_{\star,\chi}}  =  0
\end{eqnarray*}
The first equation is the dispersion relationship (\ref{drel}) while the second
is the standard Aronson-Weinberger criterium stating that the selected velocity
is the minimum velocity, a velocity that corresponds to the marginally stable solution (Eq. (\ref{awc})).

Notice that previous argument assumes that both $F(ik)$ and $A(k)$ are analytical functions.
If the initial condition decreases exponentially, i.e., $A(k)$ has a complex pole which imaginary
part is $\mu$ (as in Eq. (\ref{inicond})), then, as discussed in \cite{VS03}, $q_{\star,\chi}$ must
be replaced by $\min(q_{\star,\chi},\mu)$. This yields Eq. (\ref{qstarchi}).
As far as $F(ik)$ is concerned, it is easy to show that 
it is analytical in a strip around the real axis
$\{ z, \Im(z) \leq \mu \}$.

Let us finally remark that a general solution traveling 
at velocity $v$ of Eq. (\ref{linearT}) can be written as:
\[
Q(z_m) = A_1 e^{-q_1 z_m} + A_2 e^{-q_2 z_m}
\]
where $z_m = z_0+vm$ and $q_1$ and $q_2$ are the two complex conjugated roots
of the dispersion relation (\ref{drel}). When $v=v(q_{\star,\chi})$, 
the selected minimum velocity, 
these two roots merge and generically the solution
behaves has \cite{VS03,Brunet,CarLeDou01}:
\[
  Q(z_m) = (A z_m +B)e^{-q_{\star,\chi} z_m}
\]
We see that the asymptotic shape of the front is not precisely
Gumbel but has a subdominant correction factor.

Notice that the matching of the previous functional shape and
the shape obtained in the previous saddle point analysis, 
when one accounts for the Gaussian corrections in the integration around the saddle point, leads to famous Bramson logarithmic correction to 
the front velocity \cite{VS03,Brunet,CarLeDou01}:
\[
  x_m = m v(q_{\star,\chi}) - \frac{3}{2 q_{\star,\chi}} \ln(m)
\] 
This ``universal'' logarithmic correction to the velocity is well known for solutions of KPP equation.
A specific analysis of these finite-size corrections to scaling is beyond the scope
of this paper and will be reported in a forthcoming study. 

\section{Scaling range for tail estimation}
\label{sec-last}

One question that naturally arises from the analysis made in section 
\ref{sec_tail} is
the question of stability of $\hat{\mu}(\nu,\chi) = q_{\nu,\chi}$ as a function of $\nu$ 
(and thus of $k$) and $\chi$. 
This question is linked to the question
of the scaling range 
associated with the log-log representation of Eq. (\ref{texpp}). 
This problem is important for pratictal purpose.
Let $p = \Delta q / q$ be the precision above which one can detect
a tail exponent variation (for the sake of simplicity we will consider
$p  \simeq 0.1$)
Let $\ln(S)$ be the scaling range over which one observes
the power-law. According to our description, this corresponds
to a variation of the H\"older exponent, i.e., 
\[
  \ln(S) \simeq \Delta \alpha \ln(T/\tau)
\]
This variation corresponds therefore to a variation of $q$ that
is
\[
  \Delta q \simeq \Delta \alpha \frac{\partial q}{\partial h(q)} =
  \frac{\Delta \alpha}{\zeta''(q)|_{q=q_{\nu,\chi}}}
\]
and finally, if the scaling range is such that $\Delta q = p q$,
one obtains
\begin{equation}
\label{srange}
  \ln(S) = p q_{\nu,\chi} \ln(T/\tau) \zeta''(q_{\nu,\chi})
\end{equation}
We see that this scaling range is related to the global scaling
range rescaled by the factor that involves the intermittency
coefficient evaluated at the value $q=q_{\nu,\chi}$.

\bibliography{ExtMuzyBacKoz}

\end{document}